\documentclass[fleqn,usenatbib]{mnras}

\usepackage{newtxtext,newtxmath}

\usepackage[T1]{fontenc}

\DeclareRobustCommand{\VAN}[3]{#2}
\let\VANthebibliography\thebibliography
\def\thebibliography{\DeclareRobustCommand{\VAN}[3]{##3}\VANthebibliography}

\usepackage{graphicx}	
\usepackage{amsmath}	

\title[Image augmentations in ML models]{The effects of image augmentations when training machine learning models in astronomy}

\author[L. Butterworth and A. Spindler]{
Leon H. Butterworth$^{1}$\thanks{E-mail: l.h.butterworth@herts.ac.uk}
and Ashley Spindler$^{1}$
\\
$^{1}$Centre for Astrophysics Research, Department of Physics, Astronomy \& Mathematics, University of Hertfordshire, College Lane, Hatfield, AL10 9AB, UK\\
}

\date{Accepted 2026 April 20. Received 2026 April 20; in original form 2026 January 29}

\pubyear{\the\year{}}

\begin{document}
\label{firstpage}
\pagerange{\pageref{firstpage}--\pageref{lastpage}}
\maketitle

\begin{abstract}
We measure the influence of image augmentations and training dataset size when training a deep neural network to classify galaxy morphology. Data augmentation is an integral step when training machine learning models and often astronomers add augmentations assuming they will always improve the performance of their models. We train multiple versions of the same pre-existing Zoobot model using different image augmentations and different dataset sizes from 230,000 galaxy images from Galaxy Zoo DECaLS to determine whether this assumption is necessarily true. We find that generally, the addition of image augmentations does improve a deep neural network's performance, however, this improvement is significantly diminished as the training dataset size increases. The choice of specific augmentations (provided they are sensible) does not seem to be as important as simply having augmentations as different augmentations result in similar increases in performances. We find that for a model of a given size, there exists a saturation point (when the model's capacity has been filled with data) that cannot be surpassed with data augmentations. We find that more complex augmentations result in longer training times and might not lead to improved performance. If augmentations are added to the training process (which is recommended), simpler augmentations might be sufficient, depending on the size of the dataset and model. We therefore encourage astronomers to carefully consider their use of image augmentations in an effort to reduce wasted time and computational resources.
\end{abstract}




\section{Introduction}
\label{s.introduction}

The use of machine learning (ML) in astronomy (and many other fields) has grown significantly over the past decade \citep{Rod22}. ML models provide a unique and efficient method for solving many problems found within astronomy, finding new solutions or optimising existing methods. Examples include detecting objects with light curves \citep{Dat19, Yu21}, object classification \citep{Che21, wal22a, jin2022}, anomaly detection \citep{wal22b, des2024}, denoising \citep{Elh21, Liu25}, dimensionality reduction \citep{Don13, Lou24} and mock image generation \citep{Spi20}.

With it's relatively low entry barrier, and wide application, ML models have already proven to be valuable tools for astronomers \citep{Smi23, zha24}. Therefore, the prevalence of ML models in astronomy will remain significant for the near future and likely into the distant future as well as surveys increase the size and scope of their missions. For instance, Euclid aims to detect 1.5 billion galaxies over six years of operation \citep{Euc25b}. As a result, Euclid will receive a truly large amount of data, to the extent that processing and utilising all available data becomes difficult due to the large volume of data, machine learning can aid in this process. For example, extracting physical properties from galaxy images \citep{Kov25}, or detecting interesting objects such as strong gravitational lenses \citep{Euc2025a} and asteroid streaks in images \citep{Pon23}. The machine learning models used are able to produce better results in faster times than previous (non machine learning based) methods. However, training and operating ML models can demand a high cost in computational resources, which can be costly to the environment and detrimental to those with limited resources and time. With ML model usage growing rapidly, it is becoming increasingly important to establish methods of limiting resource costs whilst still achieving the results ML models are already producing.

The size of the dataset used to train the model affects the learning process of ML models. For a ML model to learn effectively, the model needs to "see" a sufficient amount of data to learn general features from the data. If the model "sees" too few examples or too many examples from the training data, the model will underfit or overfit respectively \citep{Bre84, Kov95}. Generally, increasing the amount of data in the training dataset leads to increased performance from the model \citep{Wal24}. However, collecting new data might be difficult or impossible in certain situations (especially in astronomy). In those situations, data augmentations become a useful tool when training ML models \citep{Nig00, Mah22}.

Data augmentation is a common step throughout machine learning models, regardless of model or data type. This step involves the training data being randomly augmented at training time in some way (either once or through a sequence of transformations). Because of this, data augmentation allows you to artificially increase the size of the training data by creating new variations of the existing data, effectively increasing the size of the training data. This step increases the diversity and size of the training dataset which often increases the ability of the ML model to learn and extract general features of the training data, reducing the likelihood of underfitting or overfitting \citep{Die15, gon18, Gar22}. This often results in the models performing better. 

The type of transformations that are applied during the data augmentation process is reliant on the type of training data and purpose of the ML model being trained. Examples include adding random noise to the data, replacing or removing parts of the data or stretching or rotating an image. The amount of training data that is required to allow the model to learn effectively is not known before the training procedure begins and it is dependant on the model architecture. It can be helpful to think about the complexity of a ML model in terms of "Model Capacity", which is a shorthand way of describing how many functions a model is capable of learning. For example, neural networks with more layers can be considered to have a higher capacity than those with fewer layers, they are capable of learning more, and more complex, functions to map the input data to the target outputs. A model with higher capacity necessarily requires more training data in order to reach a solution, either in the form of more input samples, or more training epochs. Data augmentation can then be thought about as a way to compensate for training a higher capacity model with a small dataset.

Galaxy morphology (it's shape and structure) is strongly correlated with galactic properties and evolution such as the position and rate of star formation and quenching \citep[e.g.][]{Bai17, Spi18, Ger21}, or tidal features being reliable tracer for mergers \citep{Kaw06, Ren23}. Because of this, galaxy morphology is a useful tracer of galaxy evolution. Therefore studying galaxy morphology, over a large selection of galaxies is important as it allows us to study how galaxies and their properties evolve as a whole \citep[e.g.][]{Cam22, Ren23, Sch14}.

Galaxy morphology can be measured using many different approaches, e.g. via s\'ersic profiles \citep[e.g.][]{Vik15} which measures how light is distributed throughout the galaxy, or by using the "CAS" method \citep{Con03} which takes three parameters (concentration, asymmetry and smoothness) and combines the results to ascertain galaxy morphology. Some papers combine the `CAS' method with other methods \citep[e.g.][]{Tar18, Bau23}. Perhaps the most direct approach is to measure the morphology through visual inspection \citep[e.g.][]{Kav14}, i.e. by manually looking at an image and identifying and classifying each galaxy's morphology one at a time. Visual classification is time consuming, the process is simply too slow to be viable for large samples of galaxies. Current surveys such as the Sloan Digital Sky Survey \citep[SDSS,][]{Alm23} and the Dark Energy Spectroscopic Instrument \citep[DESI,][]{Aba22} can detect millions of galaxies. This means that it would be very difficult and time consuming to accurately classify the morphology of all of the galaxies from current surveys.

This issue led to the rise of citizen science projects such as Galaxy Zoo \citep[GZ,][]{Lin08}. The aim of citizen science projects is to get the general public to volunteer and sort through data that scientists do not have the time to sort through themselves. For instance, Galaxy Zoo DECaLS \citep[GZD,][]{wal22a} shows volunteers a picture of a galaxy and asks them to classify the galaxy's morphology by asking them to answer questions related to what morphological features they can or cannot see. Opening up the classification process to volunteers from the general public (as opposed to keeping it among a small group of astronomers) means that there are many more people who can now classify galaxy morphology, greatly increasing the speed at which morphological classifications can be obtained for large samples of galaxies. 

However, current generation surveys are starting to detect so many objects that even citizen science projects can take years to complete \citep{Mas21, wal22a}. Future surveys such as the Legacy Survey of Space and Time (LSST) are estimated to observe 20 billion galaxies \citep{Ive19}, Euclid aims to detect 1.5 billion galaxies \citep{Euc25b}, the Roman telescope is predicted to detect hundreds of millions of galaxies \citep{Mos20}. Current surveys such as DESI have already detected over 13 million galaxies \citep{des26}. Classifying the morphology of that many galaxies would take far too long, even for citizen science projects. This has lead some to turn towards machine learning as an alternative \citep[e.g.][]{Wal19, Cav23, Bau23}. ML can achieve in a matter of hours or days what would take humans months or years, this significant increase in speed is what drew researchers towards machine learning.

ML can also be used to influence the training process itself, for instance \citet{wal22a} use active learning to influence which galaxies are shown to volunteers. Active learning measures how informative a galaxy would be for the purposes of training their machine learning model. Galaxies that are believed to be more informative are therefore shown to more volunteers so that more volunteer classifications are obtained for the more informative galaxy. 

The aim of this paper is to measure the affects of image augmentations and training dataset size for a ML model of a given size, and see how that is related to the model's capacity. We use the ML model Zoobot \citep{wal22a} which classifies the morphology of galaxies based on their images. This model will be trained with different image augmentations and training dataset sizes and the accuracy of the resulting models will be compared.

Section \ref{s.data} describes the DECaLS survey data as well as the morphological classifications from GZD. Section \ref{s.zoobot} details the Zoobot code, the image augmentations that will be tested and our methodology for testing their importance with the different training dataset sizes. Section \ref{s.analysis} discusses the results and analysis and section \ref{s.conclusion} summarises the findings of this work.


\section{The data}
\label{s.data}

\subsection{Galaxy Zoo classifications}

\begin{figure}
	\includegraphics[width=\columnwidth]{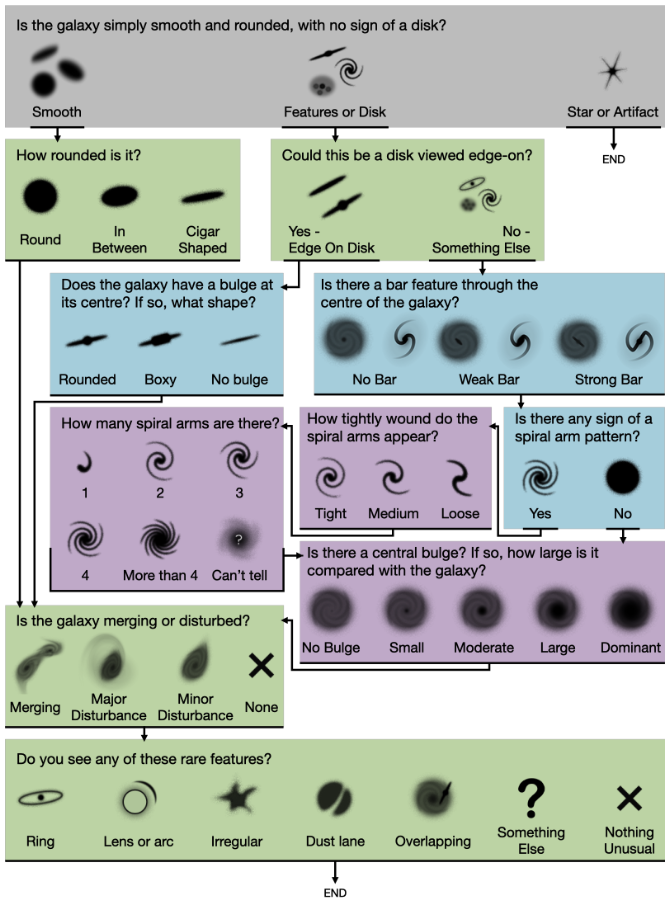}
    \caption{Decision tree used by GZD-5 \citep{wal22a}. This shows the order in which questions are shown to volunteers based on their answers to previous questions.}
    \label{f.zoobot questions}
\end{figure}

To achieve the morphological classifications for the GZD galaxies, volunteers were shown a galaxy image and asked to answer a set of morphology based questions. The set of questions form a decision tree. Figure \ref{f.zoobot questions} shows the decision tree used for GZD-5 (the fifth data release of DECaLS survey). The questions for GZD-5 were different from GZD-1 and GZD-2 (the first and second DECaLS data releases respectively, GZD-1 and GZD-2 shared the same decision tree). GZD-5 had an improved and clearer set of questions that were also designed to better classify mergers and weak bars that could now been seen in the GZD-5 data. The improvements in GZD-5 consisted (mainly) of changing either the image or name of an answer to make the answer clearer and/or more distinct from the other answers. 

Because GZD-5 has the best decision tree (as well as the most classifications), GZD-5 will be the dataset that will be used to train our models (similarly to \citealt{wal22a, Wal24}). GZD-5 has 11 possible questions volunteers could be asked. Zoobot only learns from 10 of the 11 questions, ignoring the last question regarding rare features as that is a multiple choice question and Zoobot is designed to learn from questions with only one given answer. Zoobot therefore learns from the other 10 questions. 

Most galaxies were shown to at least 5 volunteers, the active learning process (discussed in section \ref{s.introduction}) promotes certain galaxies during the classification process, resulting in these galaxies receiving over 30 votes as they are shown to more volunteers. These votes are used to assign morphological classifications to galaxies with the answer that has the most votes being taken as the "correct" answer. This has the additional benefit that the fraction of votes each answer receives acts as a confidence measurement for each classification answer. Not all galaxies can receive over 30 volunteer votes as doing so would extend the project by several years \citep{wal22a}. All images do receive multiple classifications, sufficiently many to ensure that the morphological results from said classifications should accurately represent the morphology shown in the image.

\subsection{The DECaLS survey}

All of the galaxy morphology information used in this paper is obtained from optical images taken from the the Dark Energy Camera Legacy Survey \citep[DECaLS,][]{Dey19}. DECaLS uses the Dark Energy Camera \citep[DECam,][]{Fla15} in Chile. The DECaLS survey contains both optical and mid-infrared bands, the images used by Galaxy Zoo are constructed from the \textit{grz} optical bands. DECaLS has had several data releases and GZD has had three projects that used data from different data releases, creating morphological classifications for each release.

GZD galaxies were selected from the DECaLS survey after cuts were applied to the data. The cuts were at a minimum brightness of m$_{\text{r}} = 17.77$, a maximum redshift of $z = 0.15$ and a minimum size of a Petrosian radius of 3 arcsec. The Petrosian radius is the radius at which the ratio of the surface brightness at that radius and the mean surface brightness within that radius is equal to a specific fraction  \citep{Pet76}, 0.2 is commonly used \citep{Tar18}. The first two cuts were chosen as a result of the limitations of the 8$^{\text{th}}$ Sloan Digital Sky Survey (SDSS) data release \citep{Aih11} and the last cut was chosen to ensure that all galaxies were large enough to show meaningful morphological characteristics. The magnitude cut is chosen due to the SDSS spectroscopic target selection limit. The size cut ensures galaxies have sufficient detail to accurately classify their morphology. These cuts result in 230,000 images, all of which contain classifications from GZD-5. These 230,000 images will be used to train the models in this paper.

The DECaLS images shown to volunteers are $424 \times 424$ pixels. DECaLS has a pixel scale of 0.262 arcsec per pixel. Each galaxy image was downloaded with this pixel scale, potentially with more than 424 pixels per side (up to 512 pixels per side). This was done as the pixel scale was then interpolated using equation \ref{e.sclaing} with $s$ being the interpolated pixel scale and $p_{50}$ and $p_{90}$ being the Petrosian 50 percent and 90 percent-light radius respectively. This rescaling ensures that each galaxy fits clearly within the final $424 \times 424$ pixel image.

\begin{equation}
    s = \text{max} \left( \text{min} \left( 0.04 p_{50}, 0.02 p_{90} \right),0.1 \right)
    \label{e.sclaing}
\end{equation}

\begin{figure}
	\includegraphics[width=\columnwidth]{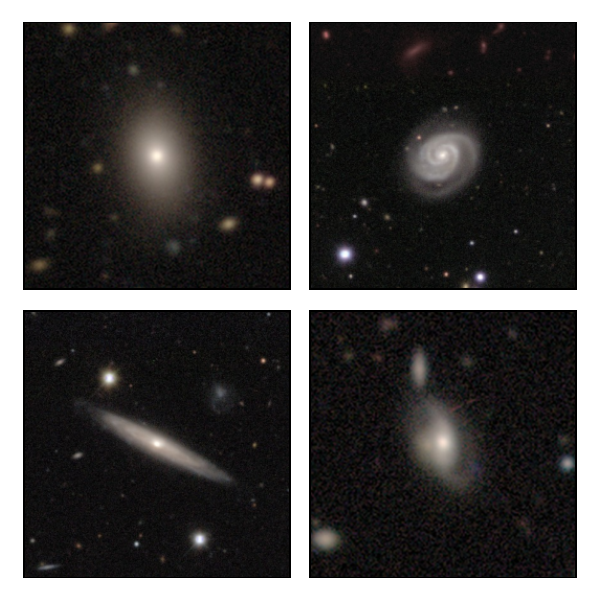}
    \caption{Selection of images from the DECaLS survey, all images were included in GZD-5. The images are created from the \textit{grz} bands. The pixel scale of each image is interpolated from DECaLS' 0.262 arcsec per pixel scale.}
    \label{f.decals images}
\end{figure}

The \textit{grz} optical band images from DECaLS are converted to RGB images by following the method in \citet{Lup04}. The fluxes in each band are multiplied by a constant value, low flux pixels are desaturated to remove colourful speckles that can arise in low flux background pixels \citep{Wil16}. The images are then scaled by $\text{arcsinh} \left( x \right)$ as this scales the images in a manner that allows for both bright and faint features to be visible without overexposing or underexposing the image. To further stop over or underexposure, the brightest and faintest pixels were capped in their brightness or dimness (having their pixel value set to a default brightness if their original value was above or below a threshold), allowing the images to occupy a range of brightness that allows for morphological features, both faint and bright, to be seen clearly. Figure \ref{f.decals images} shows multiple examples of GZD-5 images.

\subsection{Test sample}

Similarly to \citet{wal22a}, the test sample of galaxies will consist of galaxies that have at least 34 vote counts (and are not present in the training sample). This ensures that the model is being tested against galaxies that should have greater accuracy in their classifications. From that test set a different test sample will be made that will also be used to test model performance. The new subsample will be called the `confident galaxies' because each question will only consist of galaxies whose vote fraction for that question is at least 80\%. Meaning that the `confident galaxies' consists of galaxies whose answers were strongly agreed upon by the volunteers. 

We test the models using the confident galaxies because the confident galaxies should be galaxies whose features are very strongly agreed upon, so the model should be able to perform especially well on those galaxies. Because with the whole test sample, there will be galaxies that have very ambiguous answers where morphological features were unclear or volunteers struggled to accurately classify morphology. In those cases, if humans struggled then the model will probably struggle too as the model learns from those unsure answers. There will be galaxies where even experienced astronomers would disagree, this should not be the case with the confident galaxies who should have well defined features. So by testing the models using the confident galaxies, we can see how well the model performs with galaxies that should be easier to classify. Table \ref{t.no of gal} shows a breakdown of how many test galaxies the model makes predictions for each question, there is a large spread in the number of test galaxies as some questions are simply asked more often as it relates to features more commonly seen in galaxies.

\begin{table}
    \centering
    \begin{tabular}{ccc}
    \hline
        Question & \multicolumn{2}{c}{Number of galaxies in sample}\\
        \cline{2-3}
                           & All galaxies    & Confident galaxies\\
        \hline
        Smooth Or Featured & 11367           & 3725\\
        Disk Edge On       & 3831            & 3503\\
        Has Spiral Arms    & 2929            & 2024\\
        Bar                & 2929            & 585\\
        Bulge Size         & 2929            & 237\\
        How Rounded        & 7002            & 3918\\
        Edge On Bulge      & 496             & 275\\
        Spiral Winding     & 2043            & 236\\
        Spiral Arm Count   & 2043            & 684\\
        Merging            & 9883            & 4092\\
        \hline
    \end{tabular}
    \caption{Number of galaxies in the test sample for the whole test sample (All) and the confident galaxies that are used to test each individual morphology question. All galaxies in these test samples received at least 34 votes, following the same method used by \citep{wal22a}. The number of galaxies is shown for each individual morphology questions as not every galaxy is used to test every morphology question as some questions might not be relevant for certain galaxies.}
    \label{t.no of gal}
\end{table}

\section{Zoobot and image augmentation}
\label{s.zoobot}
\subsection{Zoobot description}

Zoobot \citep{wal22a} is a convolutional deep neural network based on the EfficientNet B0 architecture from \citet{Tan19}.\footnote{This paper uses an older version of Zoobot from \citet{wal22a}, a newer paper \citep{Wal24} has since been published that contains updated versions of Zoobot. Any mention or description of Zoobot in this paper (unless stated otherwise) refers to the older version of Zoobot.} Zoobot has a series of convolution layers and the output layer produces Dirichlet-Multinomial posteriors that produce vote counts for the answers of any particular question. Zoobot also uses random dropout to help with training. The EfficientNet B0 architecture was chosen because the architecture has been optimised for learning efficiency, i.e. ability to produce maximum model performance from minimal training time.

Trained to answer 10 different questions related to galaxy morphology, Zoobot is capable of classifying and providing detailed information regarding galaxy morphology. Zoobot also allows us to measure the performance of the model with regards to different aspects of morphology. i.e. we can see how changing aspects of the code affects questions about bars differently to how it affects questions about spiral arms etc. This potential for greater insight is why it has been chosen to test how image augmentation and training dataset size affect ML model performance for a model of a given size. When training, an 80/20 train-test split is used, with the data being separated randomly. However, the split (whether a galaxy belongs to the training or test dataset) remains the same for all of our models and their training.

\subsection{Zoobot's image augmentation}
\label{subs.image augmentation}

This paper is partly concerned with how data augmentation affects the performance of a ML model of a given size. As our chosen model (Zoobot) is trained on galaxy images, this paper focusses on image augmentations specifically. We only investigate augmentations that would be beneficial when studying galaxy morphology, e.g. we look at random rotations but not extreme image deformations. Image augmentation is a vital step when training ML models as it often reduces the likelihood of overfitting while training. However, when dealing with images of galaxies (or other astronomical images), the exact importance and effects (performance and training time) of individual augmentations are still not fully understood and as a result, we want to quantify the importance and affects of image augmentation.


Zoobot's transformation process consists of two steps, the preprocessing step and the actual augmentation step. The preprocessing step consists of any transformations that do not have a random component to their transformation and are applied to every image when training the model. This step is applied to all images both during the training and prediction stages. The second step consists of all augmentations that have any level of randomness in their effect on the image output. These augmentations are only applied during training and are not applied during predictions.

Zoobot's preprocessing step contains only one augmentation. Zoobot starts with the RGB galaxy images that were shown to GZD volunteers, the preprocessing step consists of converting the image to greyscale. These next steps form the augmentation step of training the model. After preprocessing, the image is randomly rotated by some angle between -180$^{\circ}$ and 180$^{\circ}$. The image is then cropped around a random point near the centre of the image. The original RGB image is $424 \times 424$ pixels, cropping the image reduces the image to $224 \times 224$ pixels. During this transformation, the image is also randomly zoomed in by a factor of 1.25 to 1.43 and stretched in either the \textit{x} or \textit{y} direction, changing the aspect ratio of the image to be between 0.9 and 1.1, the new image is created using bilinear interpolation. The last augmentation is to flip the image horizontally, there is a 50\% chance that this last step occurs (this 50\% likelihood is the same whenever a flip transformation is mentioned in this paper).

Converting the RGB images to greyscale reduces the possibility of the model learning a colour dependence. This is desirable as colour is not a reliable proxy for morphology \citep{Sme22}. The rotation and flip augmentations both increase the diversity among the training galaxies \citep{Tay18}, creating a model that should be rotationally invariant \citep[e.g.][]{Die15}. Adding the possibility to horizontally flip an image effectively doubles the size of the training dataset while rotating the images provides, in theory, a much larger dataset as images can be rotated by many different angles. Cropping the image reduces the number of inputs from 179,776 to 50,176 pixels (inputs) per image, increasing the speed at which the model can learn. Randomly cropping around a point near the centre of the image also increases the diversity among the training sample. Zooming into and stretching the image increases the size of smaller features in the image e.g. bars or merger features, increasing the likelihood of the model learning from those features.

\subsection{Changing Zoobot's image augmentation}

\begin{figure*}
	\includegraphics[width=0.9\textwidth]{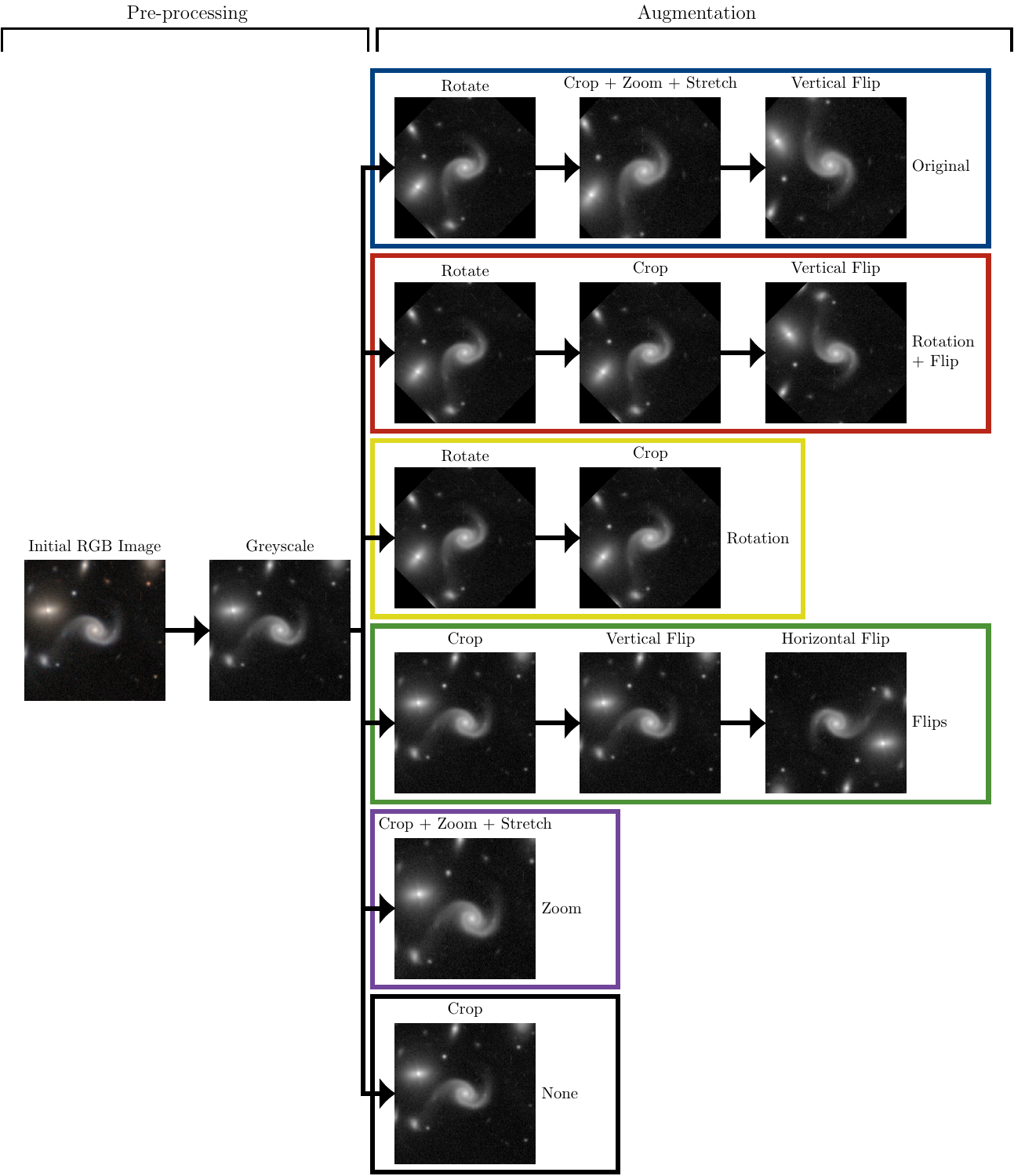}
    \caption{Visual representation of the transformation process for the six model configurations tested. Each transformation step is shown as an example image and the transformation process moves from left to right (denoted by the arrows between images). Each model configuration occupies its own row. The transformation processes contain two sections, "Pre-processing" which refers to transformations that are applied equally to every image during training, which are therefore represented as one unified row and the "Augmentation" section which displays the random transformations that are applied to each model configuration (separated by rows). The Transformation that occurs in each image is written above the image and model name is written to the right of the final image in that model configurations process.}
    \label{f.transformations}
\end{figure*}

In order to test the importance of each of these steps, the model was trained multiple times with different levels of augmentations i.e. some of the transformation steps were removed. This allows us to compare how the model performs while using the different augmentations during training. We created six different model configurations. They are called 'Original', 'Rotation + Flip', 'Rotation', 'Flips', 'Zoom' and 'None'. All six model configurations have different transformation processes. The transformation process (including preprocessing, which remains the same for every model configuration) for each model configuration is shown in figure \ref{f.transformations}. 

Figure \ref{f.transformations} shows the transformation process for our six model configurations. The configurations are separated into individual rows that show the complete transformation process. Each model configuration starts with an RGB galaxy image, an example is shown as the left most galaxy in figure \ref{f.transformations}. The subsequent images each show the effects of an individual transformation on the galaxy image, the process taking place in each image is written above the image. The transformation processes for each configuration starts with the "Initial RGB Image" and each subsequent transformation step is denoted with an arrow showing the order of transformations. 

Each transformation process consists of two sections, the "Pre-processing" and the "Augmentation" sections. The two sections are shown in figure \ref{f.transformations}. The "Pre-processing" section contains all transformations that form the pre-processing stage of training (i.e. converting the RGB image to greyscale) and the "Augmentation" section contains all other transformations, the transformations that are applied with some form of randomness different to the other model configurations. As the "Pre-processing" section contains information that is the same for each configuration, it is represented as one row that branches into the six separate rows/configurations in the "Augmentation" section. The name of each configuration is written to the right of the final image/step in each configuration.



For every model configurations, the $424 \times 424$ RGB image is the starting point of our augmentation process The first step (preprocessing) for all of our models involves converting the RGB image to greyscale. The Original model configuration consists of the original Zoobot transformation process described in section \ref{subs.image augmentation}. The Rotation + Flip model configuration is the same as the Original model configuration except for the fact that the Rotation + Flip model configuration does not zoom into or stretch the image. The Rotation model configuration is the same as the Rotation + Flip model configuration except for the fact that this model configuration does not flip the image. The Flips model configuration is similar to the Rotation model configuration, except the Flips model configuration replaces the rotation transformation with a vertical and horizontal flip, each with an independent 50\% chance to occur. The Zoom and None model configuration both only crop the image, the difference between the two models is that the Zoom model configuration zooms into, stretches and crops the image while the None model configuration does not have this additional transformation step and only crops the image.


Steps such as rotating and flipping increases the diversity among the sample, however, at some point a training set of images should become so large and naturally diverse compared to the model complexity that it removes the need to artificially increase the diversity and sample size. The same is potentially true for steps such as the cropping and zooming in transformation, if the model sees enough examples of bars and mergers etc. those features should not need to be artificially promoted as the model should see enough examples to accurately learn from them without the need for augmentations. If the training sample is naturally large and diverse enough, changing the augmentations should have very little to no effect on the performance of the model as it reaches a saturation point where augmentations do not affect the model's ability to learn.

To test if we reach the saturation point, we create three subsamples from the original $\sim$180,000 training galaxies that make up the training dataset (80\% of our 230,000 images). Those three subsamples contained 50\%, 25\% and 10\% of the original $\sim$180,000 galaxies chosen at random. Together with the original training set, that means there are four different samples of training images with different numbers of galaxy images in each for the model to train on. By training the model using those different sample sizes, we can assess how much of an impact training sample size has on Zoobot and whether we are approaching the saturation point with the DECaLS data and model size.

That means there are four different training datasets and six different model configurations with different image augmentations, resulting in 24 different models  that can be compared with each other to test how image augmentation and training sample size affect Zoobot. There are certainly other transformations that haven't been mentioned here that could also be used such as randomly changing contrast or brightness and there are also many other permutations that could be created with the transformations that have been tested here, however the tests here should be detailed and different enough to quantify the importance and impact of some of these transformations and training sizes.

\subsection{Early stopping}

All models were run on the same machine in the University of Hertfordshire's high-performance computing facility and used early stopping. Early stopping means that if the model does not improve after a certain number of epochs, the model finishes training as it has essentially stopped learning at this point. This is useful because before running the model, it is impossible to accurately guess for how long the model will keep improving and learning, because at some point the model will stop learning as it reaches its maximum capability, beyond this point the model can actually decrease in ability \citep{Yin19}. Zoobot has a patience of eight. This means that the if the model does not improve after eight epochs, the model finishes training. This is not only useful because it helps the model reach its optimum ability, but it also allows us to measure how long the model takes to learn, and at what point does it plateau. 

The model measures performance based on the loss from the validation set. The validation loss is calculated for each of the 10 morphology questions and then the sum of every questions loss is used as the overall loss value for the model. This final loss value is how the performance of the model is measured during training. It is the overall loss value that is used for the early stopping mechanism. This does mean that at the point of stopping, the model could still be improving on several questions while worsening or stagnating on other questions, leading to an overall decrease in ability despite some improvement in one or two questions. However, since the model has to answer every question as accurately as possible, this is just an unfortunate trade-off of the way the model learns and evaluates its ability.

\section{Analysis}
\label{s.analysis}
\subsection{Model performance}

\begin{figure*}
	\includegraphics[width=0.8\textwidth]{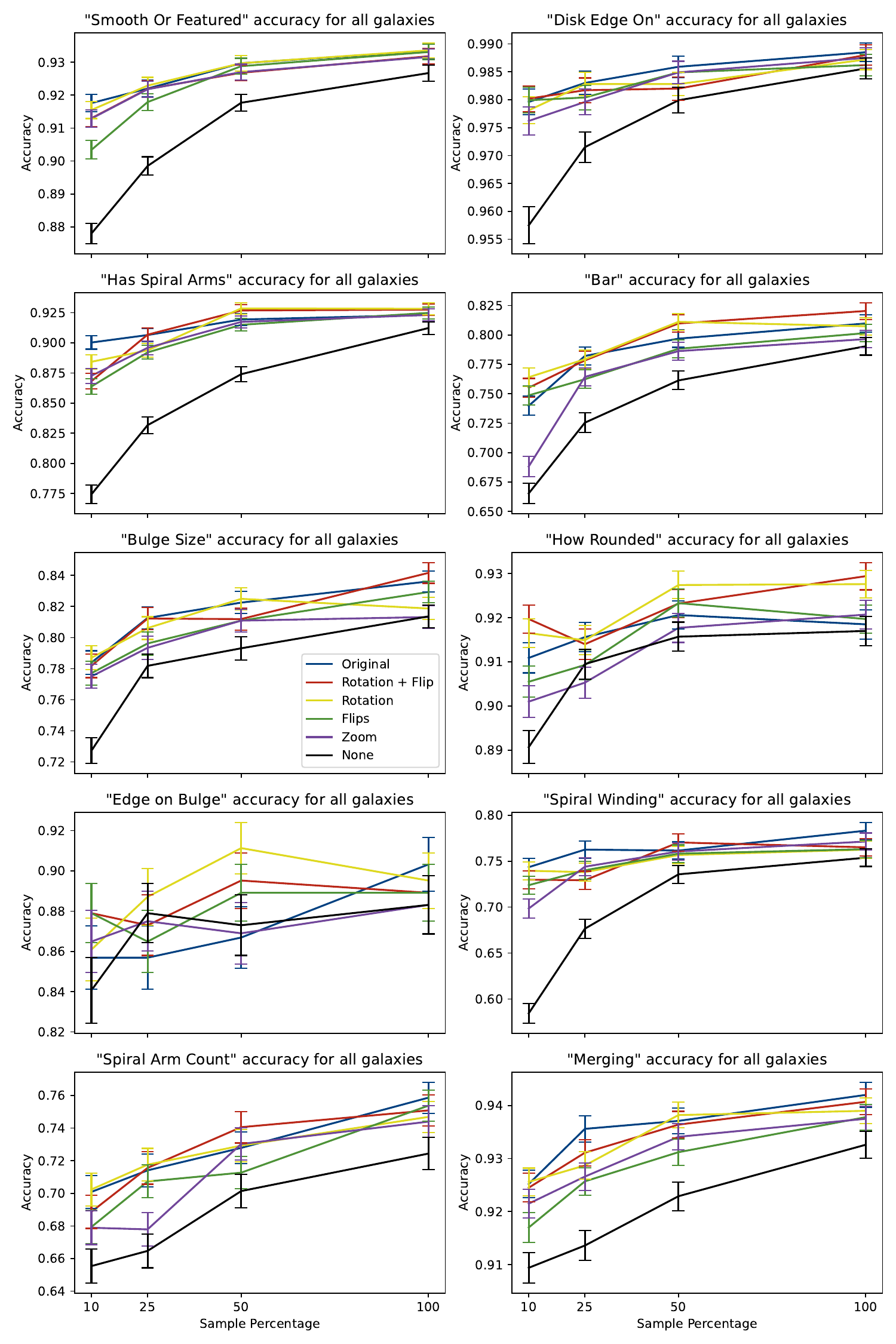}
    \caption{Accuracy for all galaxies in the test sample, the x-axis shows the size of the training sample used to train the model (as a percentage of the total training size). The question being tested in written above each plot.}
    \label{f.acc all1}
\end{figure*}

\begin{figure*}
	\includegraphics[width=0.8\textwidth]{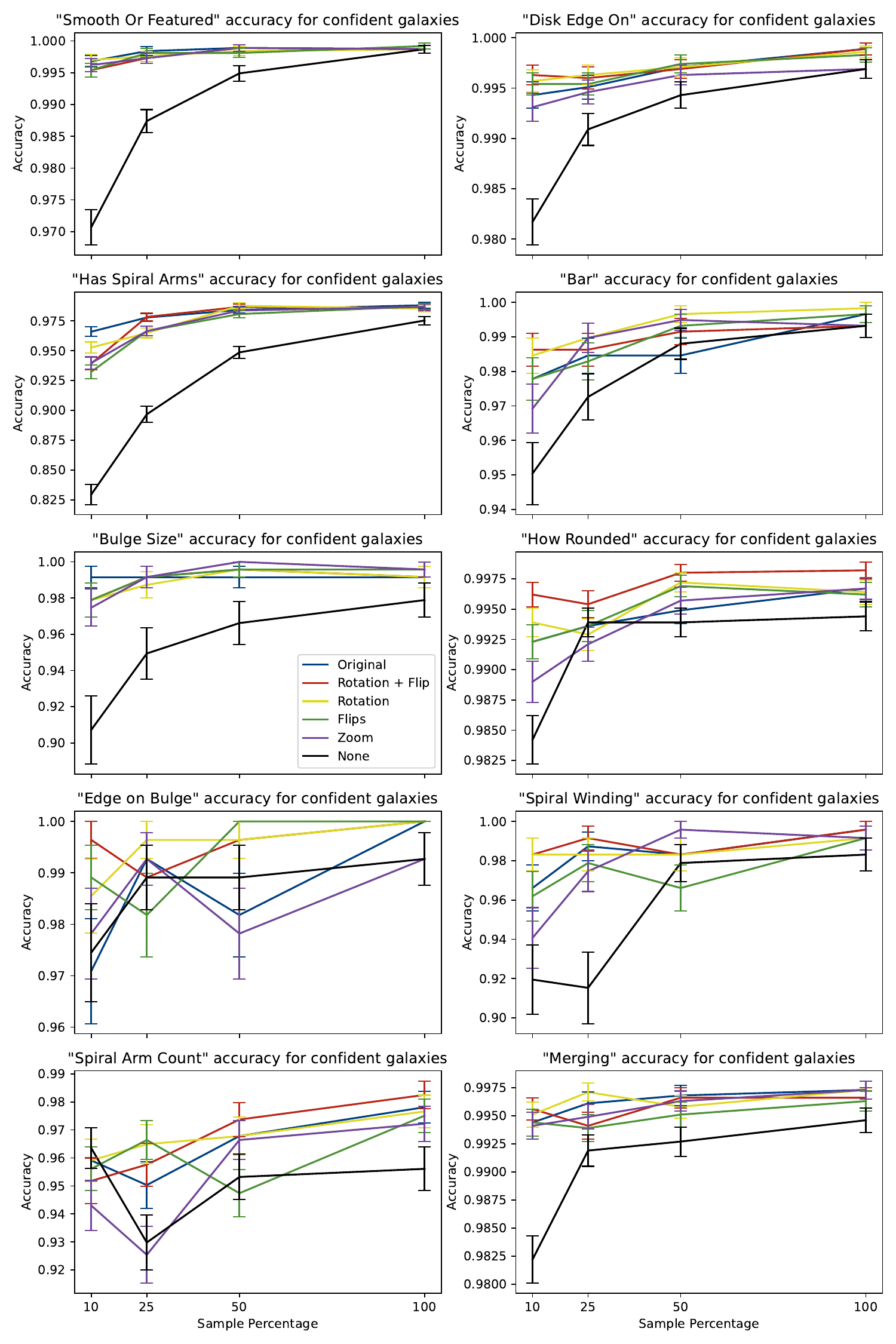}
    \caption{Accuracy for confident galaxies in the test sample, the x-axis shows the size of the training sample used to train the model (as a percentage of the total training size). The question being tested in written above each plot.}
    \label{f.acc conf1}
\end{figure*}

To measure each model configuration's performance, we will use the accuracy of the model. Accuracy is calculated as the number of correct classifications divided by the total number of classifications made by the model i.e. the fraction of correct classifications. The classifications are determined by getting the model to make predictions on a group of galaxies previously unseen by the model (a test set of images). The model then makes five forward passes with random dropout each time, the mean from the five posteriors is then used as a final classification prediction by the model. 

The model will make predictions for all 10 questions, even if the question is not suitable to the galaxy. Therefore, when evaluating the model's performance, we only consider a specific question for a galaxy if at least half of the volunteers answered that question, that shows that at least half of the volunteers agreed that that specific question was relevant to the galaxy. This allows us to avoid a situation where we are testing the model's ability on unrelated questions e.g. asking the model to predict the number of spiral arms for a galaxy that has no spiral arms. This does mean that each question was tested on a different number of galaxies.

Figures \ref{f.acc all1} and \ref{f.acc conf1} show the accuracies (for the 10 questions from the GZD-5 decision tree) of the different model configurations with different training dataset sizes for all test galaxies and the confident test galaxies respectively. All accuracies have a uncertainty calculated by jackknife sampling. \footnote{figures \ref{f.acc all} and \ref{f.acc conf} show the exact same accuracies except those plots share the same y-axis limits, meaning comparison between model configurations is easier using these figures.} Each plot in figures \ref{f.acc all1} and \ref{f.acc conf1} have the name of the question being tested above the subplot.

Figure \ref{f.acc all1} shows that between the 10 different questions, there are some common trends in how the different model configurations perform. One immediate shared trait is that the None model configuration (the model configuration with the fewest transformations) is almost universally the worst performing model configuration at every training size. This means that having transformations is beneficial as the other model configurations that do have transformations almost always perform better than the None model configuration. The biggest difference between the None model configuration and the other model configurations is often found when training with only 10\% of the training sample. The reduced number of galaxies obviously affects the None model configuration the most as the other model configurations all have augmentations that can improve the training dataset in terms of diversity and effective dataset size by "creating" new images. 

The difference between the None model configuration and the other model configurations tends to decrease as the training sample size increases. This decrease in difference between model configurations can be quite dramatic as the sample size increases, e.g. for the `Spiral Winding' question in figure \ref{f.acc all1}, the difference between model configuration accuracies decreases from over 10\% to only a few percent as the training size increases. When trained on the full 100\% of the training data, the None model configuration has reduced the difference between itself and the other model configurations meaning that it only performs slightly worse or even as well as the other model configurations. This suggests that as the size of the training sample increases, the augmentations become less important as you can achieve similar or even the same results without the transformations. 

This could be because the training size is so large and naturally diverse that augmenting the images to increase diversity and/or promote features is not necessary and only produces diminishing returns. If this is the case, it not only suggests that image augmentations might not be that necessary if you have enough data available (which we know to be true \citealt{Xu23}), but that we might be starting to reach that limit with the training data in GZD-5 and the model capacity found with Zoobot (Zoobot has 5.3 million parameters).

\subsection{Model capacity}
Another common trend in the plots in figure \ref{f.acc all1} is that the model accuracy often increases as the training sample size increases. The biggest increase often occurs between the model configurations trained using 10\% and 25\% of the training sample. This is likely because this represents the biggest change in data size comparatively in the test (i.e. a 2.5x increase compared to only a 2x increase in the other instances). However, when the model configurations go from training on 50\% to 100\% of the training sample, sometimes the model's performance increases only very slightly or even plateaus or decreases slightly. This can be seen in questions such as `Has Spiral Arms' and `How Rounded'. 

This could indicate that for these model configurations and questions, they are approaching a saturation point where the model configuration is seeing so many examples of galaxies that no image augmentations or larger training sample will improve the model significantly. While this appears to be true for the two aforementioned questions, some of the questions do still show a clear improvement between the 50\% and 100\% training sizes such as the `Spiral Arm Count' and `Merging' questions.

It is important to note that the capacity of each model configuration is exactly the same as each configuration contains the exact same architecture. The only difference between configurations is the amount/type of augmentations used in training and the size of the training data, which is consistent between configurations (i.e. the data used to train each configuration with 50\% of the data is exactly the same between configurations). Neither of these things affect the capacity of the model configurations. The capacity between questions within each configuration will be different but will be the same among configurations, e.g. the capacity of the model for the "Smooth Or Featured" questions will be different from the capacity for the "Disk Edge On" question, but both questions will have the same capacity in all of our six configurations.

The fact that some of the model configurations appear to completely fill the model capacity (i.e. reach a saturation point) for some of the questions but not for others is likely a result of two phenomena. The first being that some questions will receive more votes, therefore the model receives more training data for that question. The second reason is that some questions are more complex, for example, it is more difficult to determine the number of spiral arms than differentiating between spiral and elliptical galaxies, even for humans. This is why the model also struggles, which can bee seen by the fact that the model's accuracy when predicting spiral arm count is $\sim20\%$ lower than the accuracy when determining whether a galaxy is spiral or elliptical. Table \ref{t.vote fraction} displays the fraction of total votes each question receives (which is the same between all of our datasets), the table displays that only six of the ten questions are answered by at most 2/3 of the volunteers. So the majority of questions receive a fraction of the total votes.

Figures \ref{f.acc all1} and \ref{f.acc conf1} shows that some of our model configurations do fill their capacity entirely while training. There are methods of increasing a model's capacity, the easiest method is to increase the size of the model \citep{Wal24}. While this does work, it does not guarantee that the performance will improve and will eventually lead to the model becoming too large and complex, with a capacity that exceeds the limits of the training data size, resulting in the model overfitting to the data. All models of a fixed size will have a capacity that it cannot exceed, even with the additions of more data or augmentations.

\begin{table}
    \centering
    \begin{tabular}{cc}
    \hline
        Question & Fraction of total votes\\
        \hline
        Smooth Or Featured & 1.00\\
        Disk Edge On       & 0.77\\
        Has Spiral Arms    & 0.67\\
        Bar                & 0.67\\
        Bulge Size         & 0.67\\
        How Rounded        & 0.91\\
        Edge On Bulge      & 0.33\\
        Spiral Winding     & 0.37\\
        Spiral Arm Count   & 0.37\\
        Merging            & 1.00\\
        \hline
    \end{tabular}
    \caption{Fraction of total votes each question receives in our training dataset.}
    \label{t.vote fraction}
\end{table}


\subsection{The best model configuration}
It is clear that the None model configuration is the worst model configuration among the six model configurations tested here. However, what is not clear is if there is a model configuration that is the best performing model configuration overall. Ordering the other five model configurations in terms of ability is difficult as the performance of each model configuration varies considerably between questions. Especially as when the model configuration has been trained on the full 100\% training dataset, there is often no significant difference between the ability of the different model configurations. There are instances where some model configurations perform better than others for a particular question e.g. the Rotation + Flip and Rotation model configurations perform better than other model configurations in the `How Rounded' question in figure \ref{f.acc all1}. However, when considering all questions, no model configuration appears to be obviously or considerably better than the other model configurations (of course excluding the None model configuration which is obviously the worst performing model configuration). This is interesting as it reinforces the idea that a saturation point is being achieved for some of the questions and that no transformations or set of transformations are inherently better.

Figure \ref{f.acc conf1} shows that the confident galaxies share many properties with the whole test sample seen in figure \ref{f.acc all1}. Again, the None model configuration is often the worst performing model configuration and no model configuration makes an obvious claim as the best performing model configuration. However, there are some differences that can be seen between the confident galaxies and the whole test sample. One difference is that the model configurations perform much better for the confident galaxies than for the whole test sample. This is to be expected but it is good to confirm. One interesting feature appears to be that the the majority of the model configurations show little to no improvement when going from training on 50\% to 100\% of the training dataset. This appears to happen much more frequently in the confident galaxies than in the whole test sample. Potentially showing that because of the better confidence in results/data, more of the model configurations are reaching a saturation point as perhaps the data has less uncertainty, therefore requiring fewer galaxies to accurately train the model to full capacity.

These results lead to the conclusion that for any model, there exists a limit in performance that cannot be surpassed by the addition of more data, either through the collection of more raw data or by transforming existing data. This is seen as several questions show multiple model configurations seeming to converge upon the same accuracy, regardless of image augmentations. The difference between model configuration performance seems to be highly influenced by the training size, usually disappearing when the model configurations are trained on 100\% of the training data. Therefore, the image augmentations are not able to increase the model configurations performance beyond a certain point that varies between questions. This suggests that the performance of this Zoobot architecture is approaching its maximum ability and larger datasets or more transformations would likely offer only minute improvements at best. \citet{Wal24} found that more complex neural networks could continue to learn from larger datasets, however, those more complex model configurations should still have a maximum performance that cannot be surpassed by image augmentations. Further study could repeat the experiments in this paper with the more complex models found in \citet{Wal24}.

When only interested in one or a small number of these morphological questions, one set of transformations might result in better performance. However, when considering all 10 questions presented here, the specific augmentations do not seem to dictate performance significantly. This perhaps incentives the use minimal augmentations if available computation time is limited, and you have a sufficiently large training dataset as the shorter the transformation process, the shorter the training process. Which can be seen in table \ref{t.training time}. However, if you have a smaller training set and/or are not limited by computation time, using the original Zoobot augmentations is recommended.

\subsection{Training time}

\begin{table}
    \centering
    \begin{tabular}{ccccc}
    \hline
        Model configuration & \multicolumn{4}{c}{Total time taken to run (hh:mm)}\\
        \cline{2-5}
                 & 10\% & 25\%   & 50\%   & 100\%\\
        \hline
        Original        & 00:56  & 01:59 & 02:23 & 07:42\\
        Rotation + Flip & 00:45  & 01:28 & 02:45 & 06:47\\
        Rotation        & 00:45  & 01:26 & 02:35 & 05:22\\
        Flips           & 00:32  & 01:13 & 02:30 & 04:44\\
        Zoom            & 00:30  & 01:05 & 02:20 & 04:11\\
        None            & 00:13  & 00:36 & 01:14 & 02:51\\
        \hline
    \end{tabular}
    \caption{Time taken to complete each run for the different model configurations and training sample sizes. All model configurations were trained using a Tesla V100S.}
    \label{t.training time}
\end{table}

\begin{table}
    \centering
    \begin{tabular}{ccccc}
    \hline
        Model configuration & \multicolumn{4}{c}{Mean time per epoch (mm:ss)}\\
        \cline{2-5}
                 & 10\% & 25\%   & 50\%   & 100\%\\
        \hline
        Original        & 00:45  & 01:46 & 03:25 & 06:54\\
        Rotation + Flip & 00:43  & 01:45 & 03:26 & 06:53\\
        Rotation        & 00:43  & 01:44 & 03:25 & 06:43\\
        Flips           & 00:41  & 01:42 & 03:20 & 06:27\\
        Zoom            & 00:42  & 01:42 & 03:20 & 06:37\\
        None       & 00:44  & 01:43 & 03:22 & 06:35\\
        \hline
    \end{tabular}
    \caption{Mean training time per epoch for the different model configurations and training sample sizes. All model configurations were trained using a Tesla V100S.}
    \label{t.time per epoch}
\end{table}

\begin{figure}
	\includegraphics[width=\columnwidth]{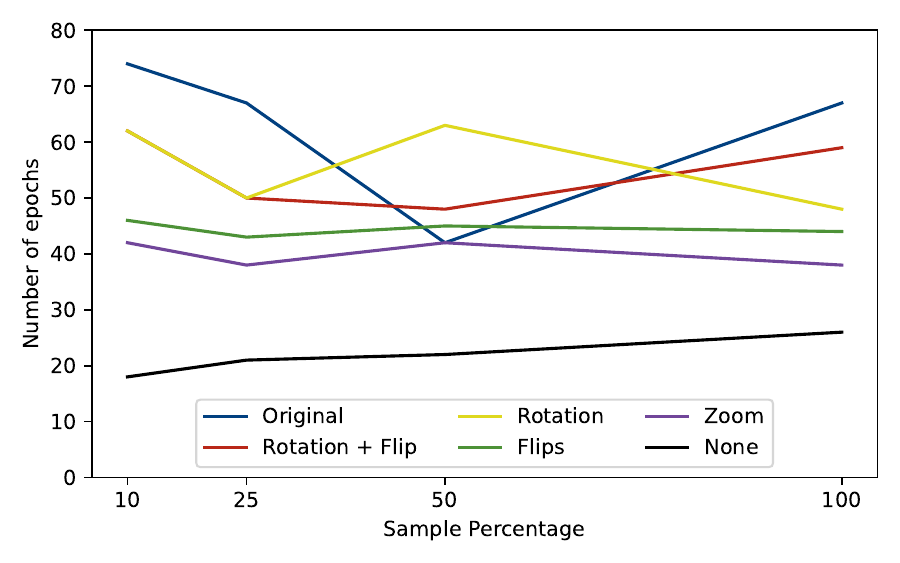}
    \caption{Number of epochs needed to train each model configuration.}
    \label{f.no of epochs}
\end{figure}

One reason why being able to determine whether our model configurations have reached their capacity useful is because training beyond that point is a waste of resources and time, both of which are limited for many. Increasing the training dataset size and number/type of image augmentations can increase the training time and resources needed to train models. Therefore understanding how the different augmentations and configurations affect the training time are important to understanding how the training times of ML models can be inflated with diminishing returns in performance in certain instances.

Table \ref{t.training time} shows the time taken to train each model configuration and the results show the expected outcome that increasing the size of the training dataset and number of augmentations increase the training time. Table \ref{t.time per epoch} shows the average time per epoch to train each model configuration. The table shows us that the the average training times per epoch do not deviate largely between model configurations that are trained on the same training dataset. The larger differences occur as the size of the training dataset increase, greatly increasing the average training time per epoch. Zoobot uses {\sc albumentations} to perform image augmentations, which is one of the fastest packages for image augmentations \citep{Bus20}. Other, slower methods of transforming images might provide larger differences in training time per epoch.

The reason why the average time per epoch is similar for each training dataset size, but the model configurations have greatly different training times in some instances, is because the number of epochs needed to train each model configuration varies greatly between model configurations. This can be seen in figure \ref{f.no of epochs}, which shows the number of epochs needed to complete each model configuration's training. From this figure, we can see that while training, most model configurations require a different number of epochs, however, each model configuration trains for a similar number of epochs even as the training dataset size increases. This means that some model configurations require more epochs to train fully and therefore take longer train.

This means that the difference in training time is related more to the number of epochs needed to train rather than the speed of the transformations themself. This means there is a simple way to predict whether a model configuration will take longer to train. If an augmentation is capable of creating more variations of the input data, the model is exposed to more images and likely require more epochs to train before it exhausts its training data. For example, flipping an image horizontally an/or vertically can increase the size of the training data by a factor of 4 essentially. Whereas rotating an image produces far more variations of the input image as the image can be rotated in more than 4 ways (our model configuration has full 360$^{\circ}$ rotation capability). This larger variation leads to more data for the model to see and therefore, results in longer training times. By extension, rotating and then flipping the image increase the diversity even more, as well as the training time. 

As we can see in figures \ref{f.acc all1} and \ref{f.acc conf1}, the difference in performance between rotating and flipping an image is often minimal or non-existent, especially when the model configurations are trained with the full training dataset. Flipping the images consistently requires fewer epochs to train and can produce similar results. The fewer number of epochs results in shorter training times. This is not to say that rotating images is inherently a waste of resources, there are instances where it produces models that outperform the Flips model configuration (e.g. the "How rounded" question in figure \ref{f.acc all1}). The conclusion of this analysis is that more does not always equal better, both in terms of training dataset size and the number/type of augmentations.

The affects of both strongly affect the time taken (and number of epochs) to train our model configurations. Our fastest model configuration trained in only 13 minutes, while our longest model configuration took almost 8 hours to train, the drastic difference in performance between these two model configurations does show the large difference in training time, with the faster model configuration being universally worse, often by a large margin. The largest gap in training time (within constant training dataset sizes) is often between the None model configuration and all other model configurations, just adding some level of augmentation greatly affects the training time (often more than doubling the training time at the least). This is reflected in figure \ref{f.no of epochs} which shows that almost all model configurations with augmentations often need two or three times the number of epochs to complete training, compared to the None model configuration. This discrepancy does not seem to affect the mean training time per epoch.

\section{Conclusion}
\label{s.conclusion}

Using data from GZD-5 we have trained the Zoobot galaxy classification model using 24 different configurations, spanning image augmentation schemes and training dataset size. Each version has one of six image transformation processes and is trained using a dataset that has one of four sizes. The ability of the different model configurations to classify galaxy morphology is then compared. We find that when the architecture of a model is fixed, there is a limit or saturation point where the model cannot improve further by adding more data to the training sample. We also find that once the performance limit has been reached, image augmentations provide inconsequential improvement and that the choice of augmentations actually provides little change in model performance when trained on large datasets. It seems that for a sufficiently large and diverse training sample, image augmentations cannot improve performance in any meaningful manner. 

The training size needed to reach this saturation point will vary between different model architectures and training data, but the saturation point should still exist for all models. This is supported by \citet{Wal24} who trained different Zoobot models on 842,000 galaxy images (far more than the 230,000 used in this paper) and found that more complex models could perform better than the more limited models who must have reached a saturation point while training. This plethora of training data could potentially mean that the augmentations used to train the newer, more complex Zoobot models could also result in diminishing returns. However, tests such as the ones performed in this paper would need to be performed in order to confirm this.

The fact that beyond a certain training size, model ability will start to plateau regardless of transformations is valuable information as including transformations increases the training time due to the increased number of epochs needed for the model to effectively "see" the new data created by the transformations. The disparity in training time can result in hours saved in training time on a computing cluster (and likely many more hours saved on a personal computer/laptop). Therefore, the ability to achieve similar results whilst saving computation time or power that might be limited is important. It seems that the full training size used in this paper is approaching the saturation point for the Zoobot architecture used in this report.

A large and diverse training sample diminishes the improvements seen by adding transformations to the training process. Importantly, when the training size is smaller than this threshold, transformations do have a strong positive influence on model ability. The smaller the training size, the larger the difference in ability between model configurations. For Zoobot specifically, it is not obvious if any particular model configuration could be classed as the best model configuration. For some questions, one or two model configurations do perform better overall than the other model configurations however considering every question results in all model configurations eventually performing to a similar standard when trained on the full training dataset. This leads us to conclude that for a model of fixed size, once a certain size and quality is reached in the training dataset, image augmentations provide little benefit and while this limit will be different for each model and dataset, this limit exists for all models. Certain transformations will generally cause the mode to require a longer training period and in situations where we might be approaching a saturation point in training, perhaps using transformations that result in quicker training times should be considered as they can likely produce similar quantity results with shorter training times.

\section*{Acknowledgements}


LHB acknowledges support from an STFC PhD studentship (grant ST/Y509218/1).  This research has made use of the University of Hertfordshire high-performance computing facility. We thank Mike Walmsley and Jim Geach for their helpful insight.

We acknowledge the relevant open source packages used in our {\sc python} codes \citep{vanros91}. {\sc albumentations} \citep{Bus20}, {\sc astropy} \citep{ast13, ast18, ast22}, {\sc numpy} \citep{Har20}, {\sc pandas} \citep{mckin10}, {\sc scikit-learn} \citep{Ped11}.

\section*{Data Availability}


The code used to test the various Zoobot model configurations is available upon request. The original DECaLS data and classifications can be found at \url{https://zenodo.org/records/4573248}.



\bibliographystyle{mnras}
\bibliography{references} 




\appendix

\section{Additional figures}

\begin{figure*}
	\includegraphics[width=0.8\textwidth]{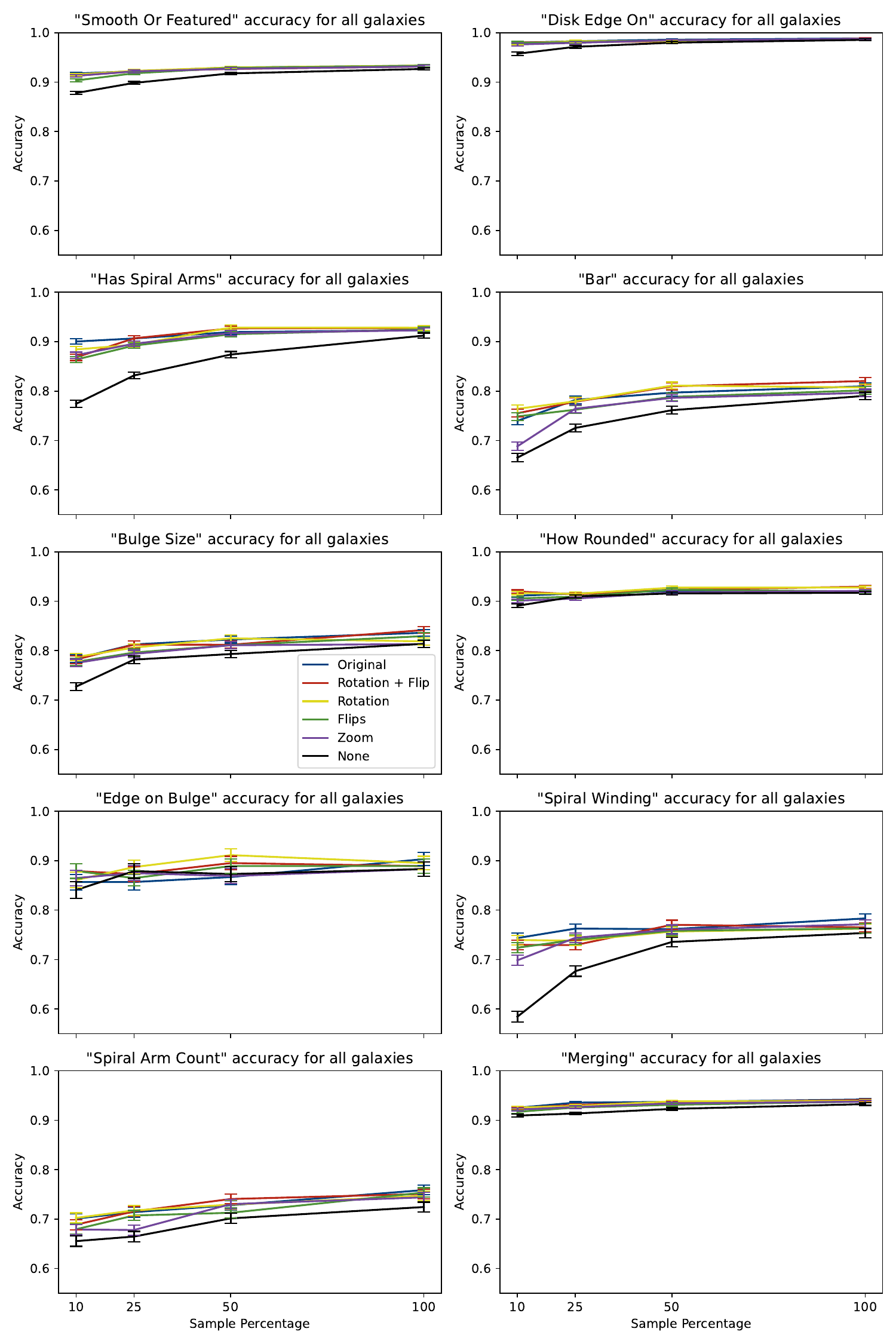}
    \caption{Accuracy for all galaxies in the test sample, the x-axis shows the size of the training sample used to train the model configuration (as a percentage of the total training size). Each plot has the same y-axis scale so that comparison between the accuracy of questions is easier. The question being tested in written above each plot.}
    \label{f.acc all}
\end{figure*}

\begin{figure*}
	\includegraphics[width=0.8\textwidth]{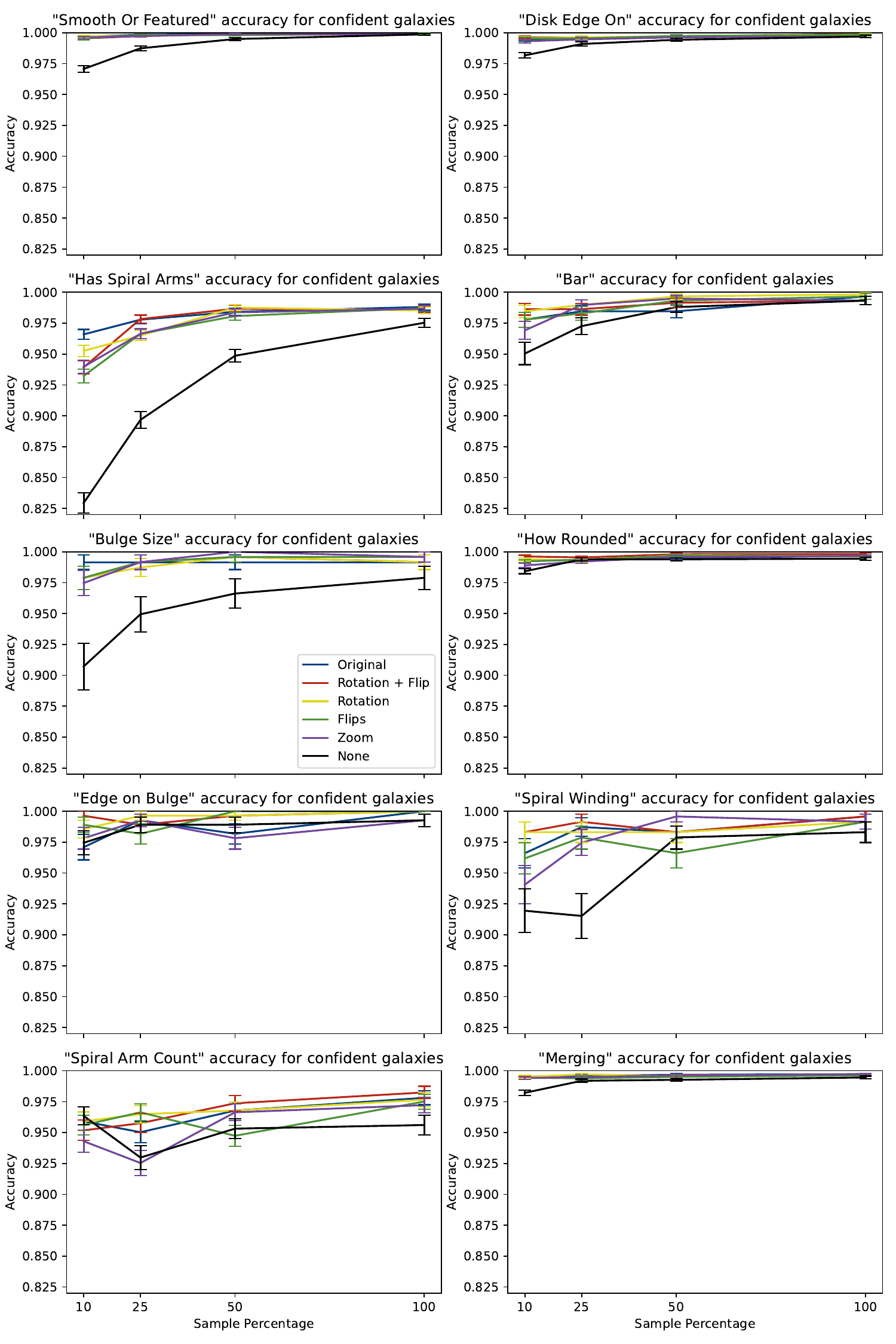}
    \caption{Accuracy for confident galaxies in the test sample, the x-axis shows the size of the training sample used to train the model configuration (as a percentage of the total training size). Each plot has the same y-axis scale so that comparison between the accuracy of questions is easier. The question being tested in written above each plot.}
    \label{f.acc conf}
\end{figure*}

Figures \ref{f.acc all} and \ref{f.acc conf} show the same data as figures \ref{f.acc all1} and \ref{f.acc conf1}. The difference between the two sets of figures is that all of the individual subplots in figures \ref{f.acc all} and \ref{f.acc conf} share the same \textit{y} axis range. allowing for easier comparison between questions.

Figure \ref{f.acc all} shows that the accuracies between questions can vary by large amounts, for instance, the `Has Spiral Arms' question has accuracies over 90\% while the `Spiral Arm Count' question has accuracies around 70\%. This means that the model is fundamentally better at replicating the answers for some questions better than other questions, by a significant amount. This could be due to the fact that the model learns from human classifications, which can lead to confusing and unsure data with large uncertainties due to the difficulty of accurately classifying galaxy morphology. Determining whether a galaxy is edge on should be simpler and easier than determining bulge size or how tightly wound the spiral arms are which leaves more room for individual interpretation (and therefore margin of error). This difference could mean that the data quality for those questions is worse and therefore leads to worse performance from the model as it is learning from this worse data.

\bsp	
\label{lastpage}
\end{document}